\documentclass[12pt]{article}
\pdfoutput=1
\usepackage{cite}
\usepackage{amssymb}
\usepackage{amsmath}
\usepackage{bm} 
\usepackage{graphicx}
\usepackage{color}
\usepackage{xcolor}
\usepackage{dsfont}
\usepackage{enumerate}
\usepackage{hyperref}
\usepackage{setspace}
\usepackage{amsmath}
\usepackage{slashed}
\usepackage[latin1]{inputenc}
\definecolor{nicered}{rgb}{0.7,0.1,0.1}
\definecolor{nicegreen}{rgb}{0.1,0.5,0.1}
\hypersetup{colorlinks,citecolor= nicegreen,linkcolor= nicered}

\newcommand{\be}  {\begin{equation}}
\newcommand{\ee}  {\end{equation}}

\def\e6{E(6)}
\def\10{SO(10)}
\def\21{SA(2) $\otimes$ U(1) }
\def\321{$\mathrm{SU(3) \otimes SU(2) \otimes U(1)}$ }

\def\422{SA(4) $\otimes$ SA(2) $\otimes$ SA(2)}
\def\roughly#1{\mathrel{\raise.3ex\hbox{$#1$\kern-.75em
      \lower1ex\hbox{$\sim$}}}} \def\lsim{\roughly}
\def\gsim{\roughly&gt;}

\def\lsim{\raise0.3ex\hbox{$\;&lt;$\kern-0.75em\raise-1.1ex\hbox{$\sim\;$}}}
\def\gsim{\raise0.3ex\hbox{$\;&gt;$\kern-0.75em\raise-1.1ex\hbox{$\sim\;$}}}
\nopagebreak
\allowdisplaybreaks
\begin{document}


\newcommand{\MP}{{\sl \small 
\sl \small $^a$ Max-Planck-Institut f\"{u}r Kernphysik, Saupfercheckweg 1, 
69117 Heidelberg, Germany\\
\sl \small $^b$ International Centre for Theoretical Physics, ICTP, 
Strada Costiera 11, 34151 Trieste, Italy\\
}}
\vspace*{0.5cm}
\begin{center}
 \textbf{\LARGE Chiral interactions, chiral states and ``chiral neutrino oscillations"} 
 \vspace{0.8cm}\\

A. Yu. Smirnov $^a, ^b$  \footnote{\href{mailto:xxxx@ictp.it}{\tt
smirnov@ictp.it}} \\
  \MP.
\end{center}
\vspace*{0.2cm}

\begin{abstract}
In vacuum the ``chiral neutrino oscillations'', {\it i.e.}  the periodic 
transitions between the left- and  
right-handed states do not occur. The produced state differs 
from the chiral component that appear in the Lagrangian of interactions 
and should be computed for each specific process.
The phase difference between components of 
a produced neutrino  is space-time independent.  This neutrino state  
consists of only positive energy solutions of the Dirac equation 
and therefore the energy splitting $2E_\nu$ between 
the  components  with different helicities 
that would drive the chiral oscillations does not exist.
Consideration in terms of neutrino propagators leads to the same conclusion. 
The situation is similar for the Majorana neutrinos and in the presence of flavor mixing. 
However, oscillations of the neutrino states produced in the chiral interactions are possible
in matter with the length determined by the matter potential. 
Description of oscillations in terms of the amplitudes of production and detection 
is elaborated that does not lead 
to any misconception. In the expanding  Universe the 
relic neutrinos adiabatically convert to 
equal number densities of the left and right handed components.   

\end{abstract}

\vspace*{0.8cm}


\newpage

\section{Introduction}

The notion  of ``chiral neutrino oscillations'' emerged  in the beginning 
of 90ties of last century \cite{FY}. It
reappeared independently few years latter 
\cite{DeLeo:1996gt} and developed over last 30 years in a number 
of publications \cite{Nishi:2005dc} - 
\cite{Morozumi:2025gmw}. 
The mass terms connect the left- and right-handed components of fermions. 
This implies some kind of transitions 
between the left and right components which according 
to \cite{DeLeo:1996gt} - \cite{Morozumi:2025gmw} have a character
of oscillations with frequency $2E_\nu$ and  depth $\approx m^2_\nu/E^2_\nu$.

The arguments go as follow. The Hamiltonian  of  the charged current 
weak interactions having the $V - A$ structure 
can be written in terms of the left handed  components of fermions as  
\begin{equation}
H \propto \bar{\nu} \gamma_\mu (1 - \gamma_5) e\, j_q^\mu  + h.c.  = 
2 \bar{\nu}_L \gamma_\mu  e_L \, j_q^\mu  + h.c., 
\label{eq:inter}
\end{equation}
where  $\nu_L \equiv P_L \nu$, $P_L \equiv (1- \gamma_5)/2$ and  
$j_q^\mu$ is the quark current. 
The key assumption in  \cite{DeLeo:1996gt} - \cite{Morozumi:2025gmw}  
is that the interactions (\ref{eq:inter}) produce the ``chiral neutrino state'' 
whose wave function immediately reflects the Lorentz structure of interactions  
(\ref{eq:inter}):
\begin{equation}
\psi_{Lh}(t = 0) =   u_{L h} \equiv  P_L u_h.  
\label{eq:chir}
\end{equation}
Here
\begin{equation}
u_h = \frac{1}{\sqrt{2E_\nu}}\left(
\begin{matrix}
\sqrt{E_\nu - hp_\nu}~ \omega_h\\
\sqrt{E_\nu + hp_\nu}~ \omega_h
\end{matrix}
\right) 
\label{eq:uspin}
\end{equation}
is the bi-spinor and  $\omega_h$ is the spinor with helicity $h/2$ 
($h \equiv (s \cdot p)/ |s||p| = \pm 1$).  
For the right handed neutrinos one has   
\begin{equation}
u_{R h} \equiv  P_R u_h 
\label{eq:chirr}
\end{equation}
with  $P_R \equiv (1 + \gamma_5)/2$.  
In what follows we will refer to the states described 
by spinors (\ref{eq:chir}) and (\ref{eq:chirr}) as to the chiral states.

It is assumed that the wave function of the chiral state, 
$\psi_{Lh}$,  (and similarly $\psi_{Rh}$) evolves as 
\begin{equation}
\psi_{Lh} (t) =  U(t) \psi_{Lh} (0), ~~~~ U(t)  = e^{-iH t},   
\label{eq:evol}
\end{equation}  
where the evolution matrix $U(t)$  is determined by 
the Hamiltonian of Dirac equation: 
\begin{equation}
H = \gamma^0 {\bf \gamma} {\bf p}_\nu + m_\nu \gamma^0 = 
{\bf \alpha}{\bf  p}_\nu + m_\nu\beta \,.    
\label{eq:ham}
\end{equation}
Here ${\bf p}_\nu$ is the   3-momentum of neutrino. Then straightforward 
computations (see details in the Appendix A) give the probabilities 
of $\nu_L \rightarrow \nu_R$ and  
$\nu_L \rightarrow \nu_L$  transitions
\begin{eqnarray}
P_{LR} & = & |\psi_{Rh} (0)^{\dagger} \psi_{Lh} (t)|^2 = \frac{m^2_\nu}{E^2_\nu} \sin^2 E_\nu t, 
\label{eq:probLR}\\
P_{LL} & = & |\psi_{Lh} (0)^{\dagger} \psi_{Lh} (t)|^2 = 1 - \frac{m_\nu^2}{E_\nu^2} \sin^2 E_\nu t.
\label{eq:probLL}
\end{eqnarray}
These probabilities have oscillatory dependence on time with  
the period $\pi/E_\nu$ and the depth ${m_\nu^2}/{E_\nu^2}$.

In \cite{Nishi:2005dc} and \cite{Bernardini:2010zba} 
apart from the first quantization approach which
leads to the chiral oscillations (\ref{eq:probLL}) also the second
quantization approach with virtual
neutrino was considered. The contributions of neutrinos and
antineutrinos as well as both signs of energy
emerge. According to \cite{Nishi:2005dc} elimination of 
neutrino or antineutrino and interference term between positive 
and negative energy components require certain
"subsidiary" condition. 
Consequently,  the chiral oscillations are absent, but it can be some zero distance transition of
the left to right components. 
The QFT consideration was also provided in \cite{Bernardini:2010zba}, 
\cite{Bittencourt:2024yxi} - \cite{Morozumi:2025gmw}
with essentially the same  results as in the first quantization approach. 
Various applications, in particular, for the relic neutrinos were explored
\cite{Bittencourt:2020xen,Ge:2020aen}.  
The chiral oscillations in matter and in the magnetic fields were considered
\cite{Li:2023iys} 
with conclusion that both the period and the depth of oscillations  are modified by
the matter potential. 

In this paper we demonstrate that  in vacuum the chiral oscillations 
(periodic transformations between the left-handed and right-handed  states) do not occur. 
The key point is that the state produced in the weak (V-A) interactions 
is not the chiral state defined in (\ref{eq:chir}). 
In fact, the state should be computed  and not taken from the Hamiltonian.  
Therefore we will distinguish  the {\it chiral states} defined in (\ref{eq:chir}) and (\ref{eq:chirr})
from the {\it states produced in the chiral interactions}.

The  produced state is shown do not develop phase difference which changes during propagation. 
The state produced in the chiral interactions can oscillate in matter or in the
magnetic fields. 
However, properties of these oscillations differ from those described in \cite{Li:2023iys}. 
We explore various aspects of propagation 
of neutrino states produced in the chiral interactions. A formalism of
oscillations (of any type) is elaborated 
in terms of the amplitudes of production and detection of neutrinos.

The paper is organized as follows.  In sect. 2 
we show that the ``chiral oscillations'' imply the presence of the negative 
energy component in  propagating neutrino and argue against this presence. 
In sect. 3 we consider neutrino states produced in
the chiral interactions and their evolution. 
The chiral states (\ref{eq:chir}),  (\ref{eq:chirr}) \cite{Nishi:2005dc} - \cite{Morozumi:2025gmw} 
are not produced in chiral interactions (sect. 4). 
Determination of the produced neutrino requires identification of  eigenstates of propagation 
$\nu_{Hi}$ and computation of  
the amplitudes of their production and detection, $A_i$. 
A general formalism of oscillations in terms of these  amplitudes is presented. 
In sect.  5 the amplitudes $A_i$ are computed explicitly and their properties studied. 
In sect.  6 we present the QFT consideration which uses  the neutrino propagators
showing  that it leads to the same result as in sect. 5.  
Effect of the flavor mixing is considered in sect. 7. 
In sect. 8 we show that  oscillations of single mass state  produced 
in the chiral interactions are possible in matter.  Sect. 9  is devoted to the Majorana neutrinos. 
Sect. 10 describes evolution of the relic neutrinos. 
The summary is  presented in sect. 11.  
In the Appendix A we provide details of  derivation of the ``chiral oscillations''. 
Connection of the ``chiral
oscillations'' and Zitterbewegung effect is discussed in 
the Appendix B.  

\section{``Chiral oscillations'' and negative energies}
 
The ``chiral oscillations'' presented in the introduction can be obtained 
in different  way that follows the description of usual flavor oscillations. 
This gives another insight into the problem. 

Recall that the  bi-spinor (\ref{eq:uspin}) is the eigenstate of the Hamiltonian (\ref{eq:ham})
with  eigenvalue $E_\nu > 0$:  $H u_h = E_\nu u_h$ and $E_\nu$  does not depend on $h$. 
So, in vacuum this state is the eigenstate of propagation.

The chiral spinor (\ref{eq:chir}) can be written as    
\begin{equation}
u_{L h} =
\sqrt{\frac{E_\nu - hp_\nu}{2E_\nu}}  
\left(
\begin{matrix}
\omega_h\\
0
\end{matrix}
\right)     =   
y_{L h}
\left(\begin{matrix}
\omega_h\\
0
\end{matrix}
\right) ,
\label{eq:uspin1}
\end{equation}
where 
\begin{equation}
y_{L h} \equiv  
\sqrt{\frac{1 - hp_\nu/E_\nu}{2}} \approx  
\left\{
\begin{matrix}
1, ~~~~~~~~~~ h = -1\\
m_\nu/2E_\nu ,~~~~h = 1
\end{matrix}
\right. .
\label{eq:pref}
\end{equation}
According to Eq. (\ref{eq:pref}) 
the spinor $u_{L +}$ with ``wrong''  helicity 
is suppressed by factor $m_\nu/2E_\nu $, while $u_{L -}$ is unsuppressed. 
The bi-spinor 
\begin{equation}
v_h \equiv 
\frac{1}{\sqrt{2E_\nu}}\left(
\begin{matrix}
\sqrt{E_\nu + hp_\nu}~ \omega_h\\
- \sqrt{E_\nu - hp_\nu}~ \omega_h
\end{matrix}
\right) 
\label{eq:hel}
\end{equation}
is the eigenstate of the Hamiltonian with negative energy: $H v_h = - E_\nu v_h$.  
For neutrino moving in $z$ direction the spinors equal  $\omega_+ = (1, 0)^T$ and 
$\omega_- = (0, 1)^T$.

The normalized spinor (\ref{eq:uspin1}) $(u^{h, norm}_L)^T u^{h, norm}_L = I$, 
equals 
\begin{equation}
u^{norm}_{Lh}  =
\left(
\begin{matrix}
\omega_h\\
0
\end{matrix}
\right).
\label{eq:norm}
\end{equation}
Then the produced neutrino state with the normalized spinor (\ref{eq:norm}) is 
\begin{equation}
\psi_{Lh}(0) = u_{L h}^{norm}.
\label{eq:statel}
\end{equation}

The chiral state (\ref{eq:chir}) or (\ref{eq:uspin1}) 
is not the eigenstate of $H$ but can be decomposed into the  eigenstates of $H$ as  
\begin{equation}
\psi_{Lh}(0) =  u_{Lh}  =  y_{Lh}  u_h + y_{Rh} v_h,
\label{eq:stat2}
\end{equation}
where $y_{Rh} \equiv \sqrt{(E_\nu + h p_\nu)/2E_\nu}$. 
Expansion of the chiral state (\ref{eq:stat2}) into the eigenstates of propagation 
requires the presence of  negative energy component $v_h$. 
(It is needed to obtain zero values of lower components in (\ref{eq:uspin1})) 
Since $u_h$ and   $v_h$ are the eigenstates of $H$  
with the eigenvalues $+E_\nu$ and $-E_\nu$,  the evolution of state 
(\ref{eq:stat2}) proceeds as 
\begin{equation}
\psi_{L h} (t) =  y_{Lh}   u_h e^{-iE_\nu t} + y_{Rh}  v_h e^{+iE_\nu t}. 
\label{eq:statev2}
\end{equation}
The same result can be obtained acting on $\psi_{hL}(0)$ 
(\ref{eq:stat2}) by the evolution matrix  $U(t)$  (\ref{eq:evol}): 
$\psi_L (t) = U(t) \psi_L(0)$. 
Consequently, the  amplitude of $\nu_L \rightarrow \nu_L$ transition equals 
\begin{equation}
A_{LL} = \psi_L (0)^{\dagger} \psi_L (t) = 
y_{Lh}^2  e^{-iE_\nu t}   + y_{Rh}^2  e^{iE_\nu t}.  
\label{eq:amplLL}
\end{equation}
Notice that $y_{Lh}^2 + y_{Rh}^2 = 1$. 
Using explicit expressions for $y_{Lh}$ (\ref{eq:pref})  
and $y_{Rh}$ one finds that the  probability $P_{LL} = |A_{LL}|^2$  
coincides with that in  Eq. (\ref{eq:probLL}).
Thus,  the chiral oscillations are result of interference of two components of the
propagating state with positive and negative energies $E_\nu$,   
so that  the phase difference equals $2E_\nu t$.  
Let us reiterate,  the expansion (\ref{eq:stat2}) requires   
usage of spinor
that corresponds to the negative energy solution of the Dirac equation 
and this is the origin of the  energy split. 
 
Another important feature here is that since $E_\nu > p_\nu, ~ m_\nu$,  the ``chiral oscillation'' 
length $l_{chiral} = 2\pi / 2E_\nu$ is smaller than 
the de Broglie as well as Compton wave lengths of neutrino: 
$l_{chiral} < \lambda_{dB}, ~\lambda_C$. 
Usually localization of neutrino production point is much worse 
than the de Broglie wave and therefore  the chiral oscillations would 
be averaged out over uncertainty in the neutrino production point. 
If the localization $\sim \lambda_{dB} $ is possible, the uncertainty 
in energy would be $\delta E_\nu \sim 1/\lambda_{dB} \sim E_\nu$. In these circumstances
oscillations will be averaged over the energy for  the baseline 
$L \gg \lambda_{dB}$. So,  it make no sense to talk on 
chiral oscillations with frequency $2E_\nu$. 
Furthermore, as we will see, the state (\ref{eq:statev2})  
is unphysical and cannot be created in real physical processes. 
The evolution of the state (\ref{eq:statev2}) corresponds to evolution with positive
and negative energies
simultaneously.

\section{Neutrino states produced in chiral interactions and their evolution}

Recall that the Lagrangian of the  Standard Model is formulated in terms of the chiral fermionic fields. 
The left handed  and the right handed components of fields have definite and   different EW
charges and therefore different gauge transformations. The charged current weak
interactions have $V - A$ structure. 
The Lagrangian of interactions obeys the chiral symmetry.  
If the symmetry is unbroken and therefore particles are massless,  
the chiral states are ``good'' states: 
They coincide with definite helicity states, 
and consequently, are the  eigenstates of propagation. 
In the limit of exact chiral symmetry there is no transition 
$\nu_L \leftrightarrow\nu_R$. 

However, the chiral symmetry is broken (and the 
proposed chiral oscillations turn out to be a consequence of this breaking). 
As we will see, the state produced in specific process depends not only  
on form of the fundamental interactions,  as in Eq. (\ref{eq:chir}),  
but also on features of this  process, its kinematics.
The produced state differs from the state which appears in the
Hamiltonian of interactions. 
The simplest and well known illustration of this fact is mixing 
of heavy neutral lepton $S$ with light neutrinos. 
In this case, e.g., the electron neutrino that appears in the Hamiltonian is 
$\nu_e = \cos \theta \nu_e'  + \sin \theta S$, 
where $\nu_e'$ is a combination of light mass eigenstates. 
If the mass of $S$ is bigger than the energy release 
in a given process $Q$: $m_S > Q$, then $S$ 
can not appear and the produced state is $\nu_e'$ which differs from $\nu_e$.

The state produced in  the chiral interactions should 
be computed taking into account masses and spins of particles 
accompanying neutrino. 
The problem of computations should be considered in terms of the eigenstates of neutrino propagation 
in medium  (vacuum or matter) $\nu_{H i}$, that is the eigenstates of the Hamiltonian.   
These states do not transform in the course of propagation. 
In vacuum, the eigenstates of propagation are the states with definite mass
and helicity   $\nu_{i h}$. 

To find the produced neutrino state, $\nu^P$, one should compute the amplitudes of production 
of the individual eigenstates  $\nu_{H k}$:  $A_{Hk}^P$. 
Then the produced state is the sum 
\begin{equation}
\nu^P = \frac{1}{N^P}\sum_k A_{Hk}^P \nu_{Hk}, 
\label{eq:prodstate}
\end{equation}
where the normalization factor equals 
$$
N^P = \sqrt{\sum_k  |A_{Hk}^P|^2}. 
$$ 
For different $k$ the amplitudes $A_{Hk}^P$ should be computed using the same state of particles 
(momenta, helicities, etc.) that accompany neutrino production. 
If for some $k = j$ one finds $A_{Hj}^P = 0$, the component $j$ 
will not produce the interference effects. 

Since $\nu_{Hk}$ are the eigenstates of Hamiltonian,  
their evolution is described in terms of plane waves as 
$$
\nu_{Hk}(t) = U(t) \nu_{Hk} = e^{-iE_k t} \nu_{Hk}. 
$$ 
In general, free propagation of $\nu_{Hk}$ in terms of wave packets generates the phase factor 
\begin{equation}
e^{- i \phi_{Hk}}, ~~~~ \phi_{Hk} = E_k t - p_k x, 
\label{eq:phase}
\end{equation}
where $E_k$ and $p_k$ are the effective (averaged) energy and momenta of the wave packets 
determined by the production process.  
Therefore the state  (\ref{eq:prodstate}) evolves as
\begin{equation}
\nu^P(x,t) = \frac{1}{N^P} \sum_k A_{Hk}^P  e^{- i \phi_{Hk}} \nu_{Hk}. 
\label{eq:evolstate}
\end{equation}
Similarly, the detected state determined by a given set of external particles 
of a detection process is
\begin{equation}
\nu^D = \frac{1}{N^D} \sum_k A_{Hk}^{D} \nu_{Hk},          
\label{eq:detstate}
\end{equation}
and $N^D \equiv \sqrt{\sum_k  |A_{Hk}^D|^2}$. Again for different $k$ the amplitudes 
$A_{Hk}^D$ should be computed using the same state of particles
that accompany neutrino detection. This ensured coherence in whole the process. 
Then the amplitude of probability to detect  $\nu^D$ equals 
\begin{equation}
A^{tot}  = \langle \nu^D | \nu^P(x, t) \rangle = 
\frac{1}{N^P N^D} \sum_k A_{Hk}^P A_{Hk}^{D*}  e^{- i \phi_{Hk}}. 
\label{eq:ampl2}
\end{equation}

In sect. 5 we will show that such a coherence of neutrinos with opposite helicities 
can be realized for neutrinos produced in $\beta-$ decay and detected by scattering.

The results can be presented in more compact way.
Introducing vectors of the amplitudes of the production and detection 
\begin{equation}
{\bf  A}^P \equiv (A_{H1}^P, A_{H2}^P, ... A_{Hn}^P)^T, ~~~~
{\bf  A}^D \equiv (A_{H1}^D, A_{H2}^D, ... A_{Hn}^D)^T,  
\label{eq:amplvectors}
\end{equation}
we can rewrite the total amplitude (\ref{eq:ampl2}) as 
\begin{equation}
A^{tot}  = 
\frac{1}{N^P N^D} {\bf  A}^{D \dagger} {\bf U} {\bf  A}^P, ~~~~ 
{\bf U} = diag \left(e^{-i\phi_1}, e^{-i\phi_2} ... e^{-i\phi_n}\right).  
\label{eq:ampl3}
\end{equation}

In the case of two components the probability $P = |A^{tot}|^2$ equals  
\begin{equation}
P = \frac{1}{(N^P N^D)^2} \left|A_{H1}^P A_{H1}^{D*}  +  
A_{H2}^P A_{H2}^{D*} e^{- i \Delta \phi_{H}} \right|^2,
\label{eq:prob2}
\end{equation}
where $\Delta \phi_{H} \equiv \phi_2 - \phi_1$. 
The interference of two terms in (\ref{eq:prob2}) produces oscillations 
if the phase difference $\Delta \phi_{H}$
changes  in space-time.  
The amplitudes of production and detection  can be approximately 
the same at certain conditions. The probability should  be integrated 
over the kinematic variables of external particles according to experimental conditions. 

This  formalism corresponds to the factorization limit when three processes involved:  
production, propagation and detection of neutrinos 
can be considered separately. We confirm its results in sect. 6 using the QFT description 
of all three processes by a single Feynman diagram with neutrino propagators.  

The production amplitudes are  computed as 
\begin{equation}
A_{Hk}^P = \int dx \langle \nu_{Hk} Y^P|H_{int}(x)| i \rangle. 
\label{eq:amplpro}
\end{equation}
Here $| i \rangle$ is the initial state,  $H_{int}$ is the Hamiltonian 
of interactions and $Y^P$ are the final state particles accompanying neutrino. 
$| i \rangle = | N \rangle $ ($N$ is nucleus) in the case of $\beta-$decay, 
$| i \rangle = |e N \rangle $ for the $e-$capture, {\it etc.}  
The integration proceeds over the localization region. 
(Integration over infinite space-time would give $\delta$ functions). 

We consider propagation in vacuum or in uniform medium, so that the eigenstates 
at the production and detection are the same. 
Then similarly to (\ref{eq:amplpro}) the  detection amplitudes of  $\nu_{H k}$ equal 
\begin{equation}
A_{Hk}^D = \int dx \langle f |H_{int}(x)| \nu_{Hk} Y^D \rangle. 
\label{eq:ampldet}
\end{equation}
Here $|f \rangle $ is the final state in a detection process,  and $Y^D$ is a  particle (usually nucleus) 
with which neutrino interacts in a detector. 

In what follows we apply the formalism to various situations. 

\section{No chiral oscillations in vacuum}

Let us consider first a single Dirac neutrino, e.g. $\nu_e$, with vacuum mass $m$. 
The eigenstates of propagation are 4 independent solutions of the Dirac equation 
which correspond to neutrinos $\nu_h$ and antineutrinos $\bar{\nu}_h$ 
with two helicities $h = \pm$: $\nu_+, ~\nu_-,~ \bar{\nu}_+, ~\bar{\nu}_- $. 
Therefore in general we should compute four 
amplitudes $A_h$. 
The Hamiltonian of the charged current interactions at low energies reads
\begin{equation}
H = \frac{G_F}{\sqrt{2}} \int dx \left[ \bar{\nu} \gamma^\mu (1 - \gamma_5) e(x)\, 
J_\mu^q + 
\bar{e}(x) \gamma^\mu (1 - \gamma_5) \nu(x)\, J_\mu^{q \dagger} \right], 
\label{eq:ham1}
\end{equation}
where $J_\mu^q$ is the hadron (quark) current, and  fields in the leptonic
currents are   
\begin{equation}
\nu(x) = \int \frac{d^3p_\nu}{(2\pi)^3 \sqrt{2E_\nu}}  \sum_h 
\left[a_\nu (p_\nu, h) u_h^\nu e^{+ ip_\nu x} + b_\nu^{\dagger}(p_\nu, h) v_h^\nu e^{- ip_\nu x}
\right], 
\label{eq:nufield}
\end{equation}
\begin{equation}
\bar{\nu}(x) = \int \frac{d^3p_\nu}{(2\pi)^3 \sqrt{2E_\nu}}  
\sum_h \left[a_\nu^{\dagger}(p_\nu, h) \bar{u}_h^\nu e^{-ip_\nu x} + 
b_\nu (p_\nu, h) \bar{v}_h^\nu e^{+ ip_\nu x} \right], 
\label{eq:bnufield}
\end{equation}
\begin{equation}
e(x) = \int \frac{d^3p_e}{(2\pi)^3 \sqrt{2E_e}}  \sum_h 
\left[a_e(p_e, h) u_h^e e^{+ ip_e x} + b_e^{\dagger}(p_e, h) v_h^e e^{- ip_e x}
\right], 
\label{eq:efield}
\end{equation}
\begin{equation}
\bar{e}(x) = \int \frac{d^3p_e}{(2\pi)^3 \sqrt{2E_e}}  \sum_h
\left[a_e^\dagger(p_e, h) \bar{u}_h^e e^{- ip_e x} + b_e(p_e, h) \bar{v}_h^e e^{+ ip_e x} \right].
\label{eq:befield}
\end{equation}
Here $a_\nu^{\dagger}$ and $b_\nu$ are the operators of creation 
of neutrino and annihilation of antineutrino,  
{\it etc.} Notice that creation of particle or antiparticle 
appears with phases $e^{-ipx}$, while annihilation --  with $e^{+ipx}$.   
The  Hamiltonian ({\ref{eq:ham1}) conserves the electric charge and lepton number.  

For definiteness let us consider the process of $\beta-$decay 
$N \rightarrow N' + \bar{\nu} + e^-$,  
where $N$ and $N'$ are nuclei in the initial and final states.  The amplitude of the process equals
\begin{equation}
A_k^P = \langle \bar{\nu}_k e N' |H| N \rangle \propto 
\bar{u}_h^e \gamma^\mu(1 - \gamma_5) v_h^\nu 
\int dx e^{i(p_\nu  + p_e) x} \Psi_\mu(x)
\label{eq:amnu}
\end{equation}
for antineutrino,  and $A_k^P =  0$ for neutrino. 
$\Psi_\mu(x)$ is the nuclear matrix element.   
As a consequence of the charge conservation, the presence of electron in the final state 
selects the second term in the Hamiltonian (\ref{eq:ham1}) 
which contains $\nu$, 
that is, the production of antineutrino with phase factor $e^{- ipx}$ 
common for both helicities. (See explicit computations of the amplitudes in sect. 5.) 
Therefore the produced antineutrino state is a combination of states  
with two different helicities and the same sign of energy: 
\begin{equation}
\bar{\nu}^P = \frac{1}{N^P} \left(A^P_- \bar{\nu}_{-}  +  A^P_+ \bar{\nu}_{+}\right),
\label{eq:stateexp}
\end{equation}
where $N^P \equiv \sqrt{|A^P_-|^2  + |A^P_+|^2}$. 

Repeating general consideration of sect. 3 for this case one finds the following. 
Evolution of the state (\ref{eq:stateexp}) can be obtained acting on it by the evolution matrix
$U = e^{-iHt}$  (see  eq. (\ref{eq:evol})). Since 
the states with definite helicities are the eigenstates of the Hamiltonian:
$H\bar{\nu}_{-} = E_\nu \bar{\nu}_{-}$,  
$H\bar{\nu}_{+} = E_\nu \bar{\nu}_{+}$ with the same eigenvalue,  we obtain 
$U \bar{\nu}_{-} = e^{-iE_\nu t} \bar{\nu}_{-}$ and 
$U \bar{\nu}_{+} = e^{-iE_\nu t} \bar{\nu}_{+}$.   
Consequently, the evolution of whole neutrino state is given by
\begin{equation}
\bar{\nu}^P(t) = e^{-i\phi(t)} \bar{\nu}^P,  
\label{eq:evolsr}
\end{equation}
or explicitly, 
\begin{equation}
\bar{\nu}^P(t) =  \frac{1}{N^P}\left[A^P_-\bar{\nu}_{-}  +  A^P_+ \bar{\nu}_{+} \right] e^{- i\phi}. 
\label{eq:stategenevo}
\end{equation}

This evolution can be obtained  performing continuation 
 of the amplitude (\ref{eq:amnu}) beyond the production region.  
Immediate consequence of  (\ref{eq:stategenevo}) is 
that there no time-dependent phase difference acquired between the two components, 
and consequently, the neutrino state in (\ref{eq:stategenevo}) does not oscillate in
the course of propagation. The detected state is 
\begin{equation}
\bar{\nu}^D = \frac{1}{N^D} (A^D_-\bar{\nu}_{-} + 
A^D_+ \bar{\nu}_{+}) 
\label{eq:stateexpd1}
\end{equation}
and the amplitude of whole the process of production and detection equals 
\begin{equation}
A = \langle \bar{\nu}^D | \bar{\nu}^P(t) \rangle = e^{i\phi(t)} \langle \bar{\nu}^D | \bar{\nu}^P\rangle ,  
\label{eq:totproc}
\end{equation}
where dependence on  phase factorizes. Then the probability reads 
\begin{equation}
|A|^2 =  \frac{1}{(N^P N^D)^2} \left|A^P_- A^{D*}_- 
+ A^P_+ A^{D*}_+ \right|^2.    
\label{eq:totprocmod1}
\end{equation}

We can introduce the mixing angle $\theta^P$ that describes the helicity 
content of the produced neutrino state:  
\begin{equation}
\sin \theta^P \equiv \frac{A^P_+}{N^P} = \frac{A^P_+}{\sqrt{|A^P_+|^2 +  |A^P_-|^2}}, ~~~~
\cos \theta^P \equiv \frac{A^P_-}{N^P} = \frac{A^P_-}{\sqrt{|A^P_+|^2 +  |A^P_-|^2}},  
\label{eq:helicitymp}
\end{equation}
so that   
\begin{equation}
\bar{\nu}^P = \cos \theta^P \bar{\nu}_- 
+ \sin \theta^P \bar{\nu}_+ .  
\label{eq:nuprod}
\end{equation}
Similarly the mixing parameter for the detected state, $\theta^D$, can be introduced: 
\begin{equation}
\bar{\nu}^D = \cos \theta^D \bar{\nu}_- + \sin \theta^D 
\bar{\nu}_+ . 
\label{eq:nudet}
\end{equation}
(Thus, the chiral interactions mix helicity states).
Then the probability  
(\ref{eq:totprocmod1}) appears as 
\begin{equation}
|A|^2 = |\cos \theta^P \cos \theta^D + \sin \theta^P \sin \theta^D|^2 = \cos^2
(\theta^P - \theta^D).  
\label{eq:totprocmod}
\end{equation}
If $\theta^P =  \theta^D$,  which requires that 
$A_+^D = A_+^P$, and $A_-^D = A_-^P$, then $|A|^2 = 1$. 
In this case there is no even $1 - (m_\nu/E_\nu)^2$ suppression of the probability. 
However,  if  $\theta^P \neq  \theta^D$, then $|A| < 1$: 
the mismatch of the  produced and detected states suppresses the probability. 
In any case, there is no space-time dependence of the probability 
and therefore there is no oscillations. 
The reason is an  absence of time dependent phase difference 
between the helicity states,  
but zero distance or time-independent transition effects can occur. 
Previous discussion of zero distance effect or ``initial flavor violation" was presented 
in  \cite{Nishi:2008sc}
as well as in \cite{Nishi:2005dc} and \cite{Bernardini:2010zba}. 

\section{Amplitudes and their properties}

Let us find explicit expressions for the amplitudes 
of production, $A_h^P$,  and detection, $A_h^D$, of neutrinos with 
certain helicities in the chiral interactions. As in Eq. (\ref{eq:pref}), the following
notations will be used
\begin{equation}
y^f_{L h} \equiv \sqrt{\frac{E_f - h^f p_f}{2 E_f}}, ~~~~~
y^f_{R h} \equiv \sqrt{\frac{E_f + h^f p_f}{2 E_f}}, ~~~~~ f = \nu, e, u, d.
\label{eq:yamu}
\end{equation}
For definiteness we compute the amplitude of production of antineutrino in the $\beta-$decay. 
 We focus on spinorial part responsible for difference of amplitudes 
with different neutrino helicities.
The term in (\ref{eq:amnu})  can be written as 
\begin{equation} 
u^{e \dagger}_{h} \gamma^0 \gamma^\mu (1 - \gamma_5) v^\nu_{h} \, Q_\mu , 
\label{eq:effint}
\end{equation}
where  $Q_\mu = \{Q_0, Q_1, Q_2, Q_3 \}$    
is the nuclear matrix element. 
In the chiral representation the matrix 
$(1 - \gamma_5) = 2 {\rm diag}(I, 0)$ reduces   
the 4-dimensional Lorentz structure to the 2-dimensional structure: 
$$
\gamma^0 \gamma^\mu (1 - \gamma_5) \rightarrow 2 \sigma^{\mu} = 
2 \{ I, - \vec{\sigma} \},
$$
where $\vec{\sigma}$ are the Pauli matrices. Thus,  the leptonic  part of (\ref{eq:effint}) equals
\begin{equation}
l^\mu_{h_\nu h_e} = 2 y_{L h}^e y_{R h}^\nu \, \omega^{e \dagger}_{h} \sigma^\mu \omega^{\nu}_{h}. 
\label{eq:lmu}
\end{equation}
As before we assume that neutrinos are moving in $z$ direction.  
For definiteness, we take $h^e = -1$ for electron 
which gives the main (chirality unsuppressed) 
contribution in ultra-relativistic case: 
$y_{L-}^e \approx 1$. 
Now $\omega^e_{-} = (-s^e, c^e)^T$, where $s^e \equiv \sin \theta_e/2$, 
$c^e \equiv \cos \theta_e/2$ and $\theta_e$ is the angle between 
the 3 momentum of electron and axis $z$ 
which coincides with 3-momenta of $\bar{\nu}$.  

For the helicity unsuppressed current we have 
$h^\nu  = + 1$, $y_{R+}^\nu \approx 1$,  
$\omega^\nu_{+} = (1, 0)^T$,  and consequently,  
\begin{equation}
l^\mu_{+ -} = 
2 y_{Lh}^e y_{Rh}^\nu \bar{u}^e_h \sigma^\mu v^\nu_h = 2 y_{Lh}^e
y_{Rh}^\nu (-s^e, c^e)  \sigma^\mu 
\left(\begin{matrix}
1\\
0
\end{matrix}
\right) = 2  \{-s^e, -c^e, -ic^e, s^e \}. 
\label{eq:lmuex}
\end{equation}
With this we can write the total amplitude $A_{h^\nu h^e}$ as 
\begin{equation}
A_{+ -} = 2 \left[-s^e Q_0 -c^eQ_1 - c^e iQ_2 + s^e Q_3 \right]. 
\label{eq:lmuexa}
\end{equation}

Similarly, for the helicity suppressed amplitude, 
$h^\nu  = - 1$,  with  $y_{R-}^\nu \approx m_\nu/2E_\nu$ and  
$\omega^\nu_{-} = (0, 1)^T$,   we have 
\begin{equation}
l^\mu_{--} = 2\frac{m_\nu}{2E_\nu} (-s^e, c^e)  \sigma^\mu
\left(\begin{matrix}
0\\
1
\end{matrix}
\right)  = 2 \frac{m_\nu}{2E_\nu} \{c^e, s^e, -is^e, c^e \}. 
\label{eq:lmuexs}
\end{equation}
This gives for the amplitude 
\begin{equation}
A_{- -} = 2 \frac{m_\nu}{2E_\nu} [c^e Q_0 + s^eQ_1 -  s^e i Q_2 + c^e Q_3].
\label{eq:lmuex1s}
\end{equation}

To get an idea about effect of the hadron part 
on the amplitude we compute $Q_\mu$ for nucleon transition $n \rightarrow p$. 
The Lorentz structure of the current is  
$\gamma_\mu (1 - g_A \gamma_5)$  with $g_A$ being the axial vector coupling. 
Representing $(1 - g_A \gamma_5)$ as  
$0.5 (1 + g_A) (1 - \gamma_5) + 0.5 (1 - g_A) (1 + \gamma_5)$
we compute  the matrix elements for $V - A$ and $V + A$ parts 
separately. For  non-relativistic hadrons  we have $y_L \approx y_R  \approx 1/\sqrt{2}$. 
Furthermore, we fix the hadron helicities as $h_p = h_n = -1$. 
Then the  $V - A$ part  equals
\begin{equation}
Q_\mu^{V - A} = \bar{p}\gamma_\mu(1 - \gamma_5) n = 
 \left(-s^p e^{-i\chi_p}, c^p \right)  \sigma_\mu
\left(\begin{matrix}
- s^n e^{i\chi_n}  \\
c^n
\end{matrix}
\right), 
\label{eq:Qmu2}
\end{equation}
where $\sigma_\mu = \{I, \vec{\sigma}\}$,    $\chi_p$ and $\chi_n$  are the azimuth angles of $p$
and $n$ with respect to the plane formed by of $\nu$ and $e$ momenta. 
Straightforward computations give 
\begin{eqnarray}
Q_0 (Q_3) & = & \left[s^p s^n e^{-i\chi_p + i\chi_n} + (-) c^p c^n  \right], 
\nonumber\\
Q_1 (iQ_2) & = & \left[- s^p c^n e^{-i\chi_p } - (+)  c^p s^n
e^{i\chi_n}  \right]. 
\label{eq:qexpl}
\end{eqnarray}
Inserting these expressions into Eqs.  (\ref{eq:lmuexa}) and (\ref{eq:lmuex1s}) we find 
\begin{equation}
A_{+-}^{V-A}  =  4 c^n K_p, \, \, \, \, \,   
A_{-- }^{V-A} =  4 \frac{m_\nu}{2E_\nu} e^{i\chi_n} s^n K_p ,  
\label{eq:atot2}
\end{equation}
where factor 
\begin{equation}
K_p \equiv -s^e c^p + c^e s^p e^{-i\chi_p}
\label{eq:k-fact}
\end{equation}
depends on characteristics of electron and proton and turns out to be  
the same for both amplitudes. 
The ratio of the $V-A$ amplitudes equals 
\begin{equation}
\frac{A_{-- }^{V-A}}{A_{+- }^{V-A}} = 
\frac{m_\nu}{2E_\nu} e^{i\chi_n} \tan \frac{\theta_n}{2}. 
\label{eq:atotrat}
\end{equation}
Besides the chiral suppression factor  $m_\nu/2E_\nu$, 
the ratio depends on the angle between neutrino and 
neutron and on the azimuth angle. It can be enhanced in the direction $\theta_n = \pi$ 
which corresponds to zero angular momentum of the  $\nu - n$ system. 
The ratio is zero when $\theta_n = 0$ (the total spin equals 1). 

For $V + A$ part, 
\begin{equation}
Q_\mu^{V + A} = \bar{p}\gamma_\mu(1 + \gamma_5) n = 
\omega^{p \dagger}_h  \sigma^\mu \omega^n_h,   
\label{eq:Qmu2}
\end{equation}
we obtain  $Q_0^{V + A} = Q_0^{V - A}$ and $Q_i^{V + A} = 
- Q_i^{V - A}$. 
Consequently, the $V +A$ amplitudes  equal: 
\begin{equation}
A_{+-}^{V+A}  =  - 4 s^p e^{i\chi_p} K_n, \, \, \, \, \,    
A_{--}^{V+A} =   4 \frac{m_\nu}{2E_\nu}  c^p K_n ,  
\label{eq:atot2pl}
\end{equation}
where 
\begin{equation}
K_n \equiv c^e c^n + s^e s^n e^{i\chi_n}
\label{eq:k-fact}
\end{equation}
depends on angles  $\theta_n$ and $\chi_n$  of neutron. 

The ratio of total amplitudes 
$A_{--} =  0.5 (1 + g_A) A_{--}^{V-A} + 
0.5(1 - g_A)A_{--}^{V+A}$ (and similarly $A_{+-}$)
equals
\begin{equation}
\frac{A_{-- }}{A_{+- }} = 
\frac{m_\nu}{2E_\nu} e^{i\chi_n} \tan \frac{\theta_n}{2} 
\left[
\frac{1 +\alpha ({c^p}/{s^n}) ({K_n}/{K_p}) e^{- i\chi_n}}
{1 - \alpha ({s^p}/{c^n})({K_n}/{K_p}) e^{- i\chi_p}}
\right].
\label{eq:atotrat2}
\end{equation}
Here $\alpha \equiv (1 - g_A)/(1 + g_A) \sim 0.1$. 
So,  the ratio depends on kinematic characteristic of all particles involved in the process.

For polarized 
nuclei one could in principle study the dependence of the suppression on angles.  
In fact, one should consider whole oscillation process  including production and detection 
of neutrinos and the intergrate over kinematic variables of external particles. 

Integration of probabilities over the angles $\theta_n$ or/and $\theta_p$  
can vanish the interference terms $P_{int}$ and therefore remove oscillations. 
For instance, according to (\ref{eq:atot2}) we have 
$$
P_{int} \propto \int_0^\pi d\theta_n \sin \theta_n {A_{+-}^{V-A}}^* A_{-- }^{V-A}  + h.c. = 
\int_0^\pi d\theta_n \sin^2 \theta_n \cos \theta_n = 0.
$$

The fact that in general the amplitudes for different 
neutrino helicities and the same helicities of other particles are non-zero shows that neutrinos 
with opposite helicities can mediate interactions 
between the same states (polarizations) of external particles 
and therefore produce interference between the corresponding amplitudes. This justifies
usage of formulas (\ref{eq:prodstate}), (\ref{eq:stateexp}), (\ref{eq:stateexpd1}) 
for states with different helicities.

Another example which illustrates that the produced neutrino state 
depends on kinematics of the processes  
and not only on the form of fundamental interaction is the pion decay: 
$\pi^- \rightarrow \mu^- + \bar{\nu}$. 
The leptonic part of the matrix element which follows 
from the $V - A$ charged current is reduced after application of the Dirac 
equation for massive muon and neutrino to 
\begin{equation}
\bar{\mu}\left[m_\mu (1-\gamma_5)  + {m_\nu} (1 + \gamma_5) \right] \nu.
\label{eq:pion}
\end{equation}

The amplitude which corresponds to  (\ref{eq:pion}) can be written as 
\begin{equation}
A_{h_\nu  h_\mu} = 2\left[ m_\mu (u_{ Rh }^{\mu})^T \gamma^0 u_{L h}^{\nu} + 
 m_\nu (u_{L h}^{\mu})^T \gamma^0 u_{ R h}^{\nu} \right]. 
\label{eq:pion3}
\end{equation}
Here the first term equals 
\begin{equation}
A_{h_\nu  h_\mu}^I = 2 m_\mu y^\mu_{R h_\mu} y^\nu_{R h_\nu} 
 (\omega^{\mu}_{h_\mu})^T \omega^{\nu}_{h_\nu}.  
\label{eq:pionmu}
\end{equation}
Since muon is non-relativistic, the components 
with $h = +1$ and $-1$ have comparable values of the order 1. 
We consider decay in the rest frame of pion, so that neutrino moves in $z$ direction 
and muon -- in exactly opposite direction: 
$\theta_\nu = 0$ and $\theta_\mu = \pi$.  
Then for different combinations of helicities of muon and neutrino we obtain from (\ref{eq:pionmu})  
\begin{equation}
A^I_{+ -}  = - 2 m_\mu y^\mu_{R -} = - m_\nu  
\sqrt{\frac{E_\mu - p_\mu}{2E_\mu}}. 
\label{eq:pionmu1}
\end{equation}
The amplitude with opposite helicities of $\bar{\nu}$ and $\mu$ equals 
\begin{equation}
A^I_{- +} =  \frac{m_\mu m_\nu}{E_\nu} y^\mu_{R +} = 
\frac{m_\mu m_\nu}{E_\nu}\sqrt{\frac{E_\mu + p_\mu}{2E_\mu}}. 
\label{eq:pionmu2}
\end{equation}
Two other amplitudes vanish: $A^I_{+ +} = A^I_{- -} = 0$ 
at $\theta = \pi$. 
Similarly for the second term in 
(\ref{eq:pion3}) we find  non-zero amplitudes
\begin{equation}
A^{II}_{- +} = - 2 m_\nu y^\mu_{L +}, ~~~~ 
A^{II}_{+ -} =  2 m_\nu \frac{m_\nu}{2 E_\nu} y^\mu_{L -}. 
\label{eq:pionmu3}
\end{equation}
The sum of two terms ($A^I + A^{II}$) equals 
\begin{equation}
A_{+ -} \approx - 2 m_\mu \sqrt{\frac{E_\mu - p_\mu}{2E_\mu}}
~~~~~
A_{- +} =  - 2 m_\nu \left(\sqrt{\frac{E_\mu - p_\mu}{2E_\mu}}
 + \frac{m_\mu}{2E_\nu} \sqrt{\frac{E_\mu + p_\mu}{2E_\mu}}
\right). 
\label{eq:pionmu4}
\end{equation}
Notice that in $A_{- +}$ the  contributions of two terms of  (\ref{eq:pion}) 
are comparable. 
For fixed helicity  of muon the amplitude with ``wrong helicity'' 
of neutrino is exactly zero.
The ratio of non-zero amplitudes depends  not only on 
$m_\nu /E_\nu$,  but also on mass and energy of muon.  

Here helicity of muon determines 
uniquely the  helicity of neutrino and therefore 
neutrinos with different helicities require different 
helicities of muon, and  can not be coherent. 
Since $A_{+ -} \neq 0$,  while $A_{- -} = 0$, neutrinos with $h = +1$ and $-1$ will not appear
together in the sum (\ref{eq:prodstate}).

\section{Consideration in terms of neutrino propagator}

Complete and consistent description of the processes of production 
and detection of neutrinos is given in QFT, where neutrino evolution  
is described by the  propagator.  We denote by  
$Y^P$ ($Y^D$) the external particles that participate in the neutrino production 
(detection). Let $x_1$ and $x_2$ be the 4-coordinates 
of points  of neutrino production and detection. 
Then the amplitude of process in which virtual neutrino 
propagates between the production and detection regions can be written as 
\begin{equation}
A = \int d^4 x_1 \int d^4 x_2 M^D e^{-i \mathcal{P}_D x_2 } 
\left[\int \frac{d^4 p_\nu}{(2\pi)^4} \frac{ \slashed{p}_\nu + m}{p_\nu^2 - m_\nu^2 + i\epsilon}
e^{-ip_\nu(x_2 - x_1)} \right]  M^P e^{-i \mathcal{P}_P x_1}. 
\label{eq:qftamp}
\end{equation}
Here $\mathcal{P}_D$ is the sum  of momenta of all external particles 
(but neutrino) at the detection, where momenta
of the produced particles are with minus sign. 
$\mathcal{P}_P$ is similar quantity but for the  production. 
$M^P$ and $M^D$ are the amplitudes of production and detection of neutrino 
but without neutrino factors (spinors)  
\cite{Akhmedov:2010ms}.  The quantity in the brackets is the propagator of neutrino.

For spinors defined in Eq. (\ref{eq:uspin}) without on-shell condition we find
\begin{equation}
2 p_0 \sum_h  u_h(p_\nu) \bar{u}_h(p_\nu) = \slashed{p}_\nu + \sqrt{p_0^2 - p_z^2}.
\label{eq:sumrule}
\end{equation}
As we will see,  for macroscopic baselines (distances between the production and detection regions)
the dominant contribution to the integral over $p_\nu$ in the amplitude (\ref{eq:qftamp})
follows from the pole at $p_\nu^2 =  m_\nu^2$, or $p_0^2 - p_z^2  = m_\nu^2$. The latter corresponds to on-shell
condition, {\it i.e.} to propagation of nearly real neutrino. 
In this case we can use 
\begin{equation}
\slashed{p}_\nu + m_\nu \approx 2 p_0 \sum_h  u_h(p_\nu) \bar{u}_h(p_\nu)   
\label{eq:sumrule2}
\end{equation}
with $u_h(p_\nu)$ taken on-shell. 
Inserting this expression into (\ref{eq:qftamp}) we can introduce the amplitudes   
with neutrino helicities $h$ \cite{Akhmedov:2010ms}: 
\begin{equation}
A_h \approx \int d^4 x_1 \int d^4 x_2 M^D e^{-i \mathcal{P}_D x_2 } 
\left[\int \frac{d^4 p_\nu}{(2\pi)^4} \frac{2 p_0 u_h(p_\nu) \bar{u}_h(p_\nu)}{p_\nu^2 - m_\nu^2 
+ i\epsilon} e^{-ip_\nu(x_2 - x_1)} \right]  M^P e^{-i \mathcal{P}_P x_1}. 
\label{eq:qftamph}
\end{equation}

Attaching neutrino spinors from propagator to the amplitudes $M_P$ and $M_D$ gives 
\begin{equation}
M^D u_h(p_\nu) = \sqrt{2p_0} A_h^D(p_\nu), ~~~~~~\bar{u}_h(p_\nu) M^P  = \sqrt{2p_0} A_h^P(p_\nu), 
\label{eq:qftampcom}
\end{equation}
where $A^P_h$ and $A^D_h$ are the  amplitudes  (\ref{eq:amplpro}) 
and (\ref{eq:ampldet}) with neutrinos included.

Suppose neutrino is produced in a region localized around space-time point $x_P$
and detected 
in the region around the point  $x_D$. These two regions are separated 
by macroscopic distance $L$ and their sizes  
are much smaller than $L$. 
Introducing the local coordinates  $x_2' \equiv x_2 - x_D$ and $x_1' \equiv x_1 - x_P$ 
and changing the order of integration we can rewrite 
the amplitude (\ref{eq:qftamp}) as   
\begin{equation}
A_h = \int \frac{d^4 p}{(2\pi)^4} \frac{ e^{-ip(x_D - x_P)}}{p_\nu^2 - m_\nu^2 + i\epsilon}  
\int d^4 x_1' \sqrt{2p_0} A^P_h e^{-i (\mathcal{P}_P - p)x_1'}
\int d^4 x_2'  \sqrt{2p_0} A^D_h e^{-i (\mathcal{P}_D + p)x_2'}.  
\label{eq:qftamp1}
\end{equation}
Here we used relations in (\ref{eq:qftampcom}) and omitted constant phase factor
$\exp(-i \mathcal{P}_D x_D  -i \mathcal{P}_P x_P)$. 

The integrals over $x_1'$ and  $x_2'$ give approximate 
(due to finite volume of integration) $\delta -$ 
functions, $\delta_r$,  which express an approximate energy-momentum conservation 
at production and detection (and eventually in whole the process):  
\begin{equation}
\delta_r(\mathcal{P}_P - p) = \int_{V_P} d^4 x_1' e^{-i (\mathcal{P}_P - p)x_1'}, ~~~~
\delta_r(\mathcal{P}_D - p) = \int_{V_D} d^4 x_2' e^{-i (\mathcal{P}_D - p)x_2'}.
\label{eq:2deltas}
\end{equation}
Here 
$$
\delta_r(P) = \frac{1}{2P} \sin P r .    
$$
The widths of these functions are determined by sizes of production and detection regions $r$.  
Pluging (\ref{eq:2deltas}) into (\ref{eq:qftamp1})
we obtain  
\begin{equation}
A_h = \int \frac{d^4 p_\nu}{(2\pi)^4} 
\frac{2p_0  \delta'(\mathcal{P}_P - p_\nu) \delta'(\mathcal{P}_D + p_\nu)}{p_\nu^2 - m_\nu^2 +
i\epsilon}
A^P_h A^D_h e^{-i p_0 T  + i {\bf p_\nu L}},  
\label{eq:qftamp3}
\end{equation}
where $T \equiv t_D - t_P$, ${\bf L} \equiv {\bf x}_D - {\bf x}_P$. 

Integration  over $p_0$  can be performed in the following way.  
The Feynman propagator  has two poles in the complex plane: 
$p_0 = E(p) - i\epsilon$ and $p_0 = - E(p) + i\epsilon$. 
The contour of integration should be  closed in the lower semiplane where  
${\rm Im} p_0 =  - |{\rm Im} p_0| <  0$. Along     
the closing part the factor $e^{-i p_0 T} = e^{- |{\rm Im} p_0| T}$ in (\ref{eq:qftamp3}) 
is exponentially suppressed. This contour includes the pole with positive energy. 
(The closure in the upper semiplane that would include negative energy pole 
is not possible since it corresponds  exponentially increasing factor).  
Moreover, the approximate $\delta$-functions in nominator further favor positive energy
pole. 
Thus,  the positive energy  solution is selected by  causality and energy conservation. 

Computing residual of the pole we obtain 
\begin{equation}
A_h \approx -i  \delta_r({\mathcal{P}_D - \mathcal{P}_P})   
\int \frac{d^3 p_\nu}{(2\pi)^3} 
A^P_h  A^D_h e^{-i E(p_\nu) T  + i {\bf p_\nu L}}.  
\label{eq:qftamp4}
\end{equation}
For  fixed ${\bf p}$ the integrand of (\ref{eq:qftamp4})  coincides with expression
(\ref{eq:ampl2}). 

Possible corrections to this result are suppressed by the exponential factors: 
$e^{- L/\lambda_\nu}$, where $\lambda_\nu$ is the size of neutrino wave packet \cite{Akhmedov:2010ms}. 
As an example, for $\lambda_\nu = 10^{-5}$ cm and $L = 10 $ m  
the correction is of the order $e^{- 10^8}$.

\section{Evolution in the presence of mixing}

Let us consider for simplicity the case of two flavor mixing.  
In vacuum, the  eigenstates of propagation are the mass states with certain helicity:  
$\nu_{ih}$, $i = 1, 2$ and  $h = + 1,  -1$. 
In the basis $(\nu_{1-}, \nu_{1+}, \nu_{2-}, \nu_{2+})$ the evolution matrix 
is
\begin{equation}
{\bf U}(t) = {\rm diag} \left(e^{- i\phi_1}, ~ e^{- i\phi_1}, ~ e^{- i\phi_2}, ~ 
e^{- i\phi_2}\right) , 
\label{eq:pionmu4}
\end{equation}
where after subtraction of common phase factor  
the phases equal   $\phi_i = m_i^2t/2E_\nu$. They are different for different masses 
but the same for different helicities
and fixed mass.  
The vectors of amplitudes of the initial and final states for  a given flavor of the  
charged leptons, e.g. the electron,  can be written as 
\begin{equation}
{\bf A}^P =
\left(
\begin{matrix}
\cos \theta \cos \theta_1^P\\
\cos \theta \sin \theta_1^P\\
\sin \theta \cos \theta_2^P\\
\sin \theta \sin \theta_2^P 
\end{matrix}
\right),    ~~~~
{\bf A}^D =
\left(
\begin{matrix}
\cos \theta \cos \theta_1^D\\
\cos \theta \sin \theta_1^D\\
\sin \theta \cos \theta_2^D\\
\sin \theta \sin \theta_2^D  
\end{matrix}
\right).   
\label{eq:infine}
\end{equation}
The helicity mixing angles $\theta_i^P$ and  $\theta_i^D$ are defined in 
(\ref{eq:helicitymp}) and (\ref{eq:nudet}). 
Then the amplitude of $\nu_e \rightarrow \nu_e$ 
transition (\ref{eq:ampl3})  equals
\begin{equation}
A =  {\bf A}^{D \dagger}{\bf U}(t) {\bf A}^P =
\cos^2 \theta +   e^{-i\Delta\phi}  \sin^2\theta  - 
2 \left[ \cos^2 \theta  \sin^2 \Delta \theta_1 +  
e^{-i\Delta\phi}  \sin^2\theta \sin^2 \Delta \theta_2 \right], 
\label{eq:ampfh}
\end{equation}
where $\Delta \phi \equiv \phi_2 - \phi_1$, $\Delta \theta_1 \equiv 0.5 (\theta_1^D
- \theta_1^P)$, 
$\Delta \theta_2 \equiv 0.5 (\theta_2^D - \theta_2^P)$. 
If there is no mismatch of the helicity mixing angles at the detection and production, 
$\Delta \theta_1 = \Delta \theta_2 = 0$, the term in brackets of 
Eq. (\ref{eq:ampfh})  
vanishes and we obtain the standard expression for the oscillation amplitude. 
If mismatch is the same for both mass states: 
$\Delta \theta_1 = \Delta \theta_2 = \Delta \theta$, the expression (\ref{eq:ampfh})
becomes 
\begin{equation}
A =  
\cos^2 \theta +   e^{-i\Delta\phi}  \sin^2\theta [1 -  \sin^2 \Delta \theta].    
\label{eq:ampfh1}
\end{equation}
The correction describes the zero distance (time-independent) effect: for $\Delta \phi
= 0$, 
Eq. (\ref{eq:ampfh1}) gives $A =  1 -  \sin^2\theta \sin^2 \Delta \theta$.

\section{Oscillations of states produced in the chiral interactions in matter}

Let us consider a single Dirac neutrino with mass $m_\nu$ in medium. 
The Hamiltonian (\ref{eq:ham}) in matter in the basis of helicity states  
$(\nu_-, \nu_+)^T$ equals
\begin{equation}
\hat{H} =
\left(
\begin{matrix}
m_\nu^2/2E_\nu  & 0\\
0 & m_\nu^2/2E_\nu 
\end{matrix}
\right) +  \hat{V}, 
\label{eq:hamexpl}
\end{equation}
where $\hat{V}$ is the matrix of  matter potentials given by 
\begin{equation}
V_{h, h'}^\mu =  \langle \nu_h | \bar{\nu}\gamma^\mu (1 - \gamma_5) \nu |\nu_{h'}
\rangle V_0,  ~~~~
V_0 = \frac{G_F}{\sqrt{2}} n_e B_f. 
\label{eq:potent}
\end{equation}
Here $B_f$ takes into account the elastic forward scattering of neutrinos on
all components of medium 
and it depends on flavor of neutrino state $f$. 
(Essentially this is the active  - sterile neutrino potential). 
The matrix element in (\ref{eq:potent}) with $\gamma^0$ 
(valid for unpolarized and non-relativistic matter) equals
\begin{equation}
\langle \nu_h | \bar{\nu}\gamma^0 (1 - \gamma_5) \nu |\nu_{h'} \rangle
= 2\langle \nu_h |\nu_L^\dagger \nu_L |\nu_{h'} \rangle,  
\label{eq:matelem}
\end{equation}
and explicitly
\begin{equation}
V_{h h'}^0  \propto 2 u_{L h}^\dagger u_{L h'} = 2 y_{Lh}^\nu y_{Lh'}^\nu \omega_h^T
\omega_{h'}.  
\label{eq:matelem1}
\end{equation}
If neutrinos in the initial and final  states move along the axis $z$,  then  
$\omega_+ = (1, 0)^T$, $\omega_- = (0, 1)^T$. 
Inserting these spinors in (\ref{eq:matelem1}) and (\ref{eq:potent}) we find
\begin{equation}
V_{--} = V_0, ~~~ V_{++} =  \left( \frac{m_\nu}{2E_\nu} \right)^2 V_0, ~~~ V_{-+} = V_{+-} = 0.   
\label{eq:matelem2}
\end{equation}
Vanishing off-diagonal elements is a consequence of conservation of angular momentum: 
flip of helicity is not possible when initial and final neutrinos 
are moving in the same direction (the refraction condition). Thus,  $\hat{V} = diag[1, (m_\nu/2E_\nu)^2] V_0$, and 
\begin{equation}
\hat{H} =
\left[
\begin{matrix}
m_\nu^2/2E_\nu + V_0^f (1 - (m_\nu/2E_\nu)^2) & 0\\
0 &  m_\nu^2/2E_\nu 
\end{matrix}
\right].
\label{eq:hamexpl1}
\end{equation}
The matter potential does not mix helicity states but produces the level splitting 
$$
\Delta V^f = V_0^f \left[1 - \left(\frac{m_\nu}{2E_\nu}\right)^2 \right].
$$ 
Due to this splitting the phase difference 
$\phi(t) =  \Delta V t \approx V_0 t$ is generated and the produced state evolves as 
\begin{equation}
\nu_P(t) = A^P_- \nu_-  e^{- i \Delta V^f t} +  A^P_+ \nu_+. 
\label{eq:evolsr}
\end{equation}
The detected state is $\nu^D = A^D_- \nu_- +  A^D_+ \nu_+$. 
Consequently, the amplitude of $\nu_f - \nu_f$ transition expressed in terms of the chiral mixing angles   
equals
\begin{equation}
A = \cos \theta^P \cos \theta^D  e^{- i\phi^f(t)} + \sin \theta^P \sin \theta^D.  
\label{eq:totprocmodm}
\end{equation}
It  gives  the survival probability
\begin{equation}
|A|^2 = (\cos \theta^P \cos \theta^D)^2 + 
(\sin \theta^P \sin \theta^D)^2 -  1/2   \sin 2 \theta^P \sin 2 \theta^D \cos
\phi(t). 
\label{eq:totprocmodm1}
\end{equation}
The depth of oscillations is determined by the chiral mixing 
of helicity states and is not affected by matter.


\section{Majorana neutrinos}

The issue of existence and dynamics of oscillations of Majorana
neutrinos produced in the chiral interactions are exactly the same as in
the  Dirac neutrino
case considered above. In particular, 
for single Majorana neutrino the eigenstates of propagation 
are two states with opposite  helicities. 
No oscillations occur in a single mass case since no time-dependent phase difference is generated. 
Some zero distance effect can be realized in the case 
of difference of the production and interaction amplitudes. 
The oscillations are possible in mater or magnetic field.

To compute the amplitudes of production and detection of the helicity
states and thus determine the chiral mixing} one
should use  the Majorana neutrino field
\begin{equation}
\nu(x) = \int \frac{d^3p_\nu}{(2\pi)^3 \sqrt{2E_\nu}}  \sum_h
\left[a_\nu(p_\nu, h) u_h^\nu e^{+ip_\nu x} + a_\nu^{\dagger}(p_\nu, h) v_h^\nu e^{-ip_\nu x} \right]. 
\label{eq:numaj}
\end{equation}
Again the consideration is similar to that in the Dirac case. For the beta decay 
it leads to the same spinorial structure as in 
Eq. (\ref{eq:effint}), and consequently, to the same expression in Eq. (\ref{eq:lmu}) for the leptonic current. 

The difference from the Dirac case is related to properties of the right-handed
neutrino components. In the Dirac case they are  the sterile neutrinos
while in the Majorana case they are the (active) antineutrino.
In the first case we have just suppression of signal while 
in the second case the active antineutrino appears. 
Thus, in the case of zero distance effect  
neutrino produced by electron can in turn produce positron, 
{\it i.e.}, the lepton number violation occurs similarly to the  neutrinoless double
beta decay. 

The matter potential which drives oscillations of
Majorana neutrinos differs from that in the Dirac case.
The oscillations proceed 
with violation of the lepton number such that neutrino  oscillates into antineutrino
(that is, the state which produces positrons).  

\section{Low energy limit. Relic neutrinos}

In \cite{Bittencourt:2020xen,Ge:2020aen} applications 
of the  ``chiral oscillations'' to the relic neutrinos  were
considered.  At the present epoch at least two neutrinos are non-relativistic, $E_\nu \approx m_\nu$ and
$p \approx 0$. 
For them the chiral suppression factor is absent and  
the mixing of the left- and right- handed components becomes maximal. 
Therefore according to \cite{Bittencourt:2020xen,Ge:2020aen} $\nu_L \leftrightarrow \nu_R$ oscillations 
proceed  with maximal depth leading (after averaging of oscillations) 
to equal number densities of $\nu_L$ and $\nu_R$. 

Although these chiral oscillations of relic neutrinos do not occur,  
the correct consideration produces the same result. 
Indeed, in the early Universe at decoupling the neutrinos were  ultrarelativistic and therefore 
produced  in states close to the 
helicity eigenstates: $\nu \approx \nu_-  + \alpha \nu_+$  
with $\alpha = O(m_\nu/E_\nu)$  (and $\bar{\nu} \approx \bar{\nu}_+$). They evolve as   
\begin{equation}
\nu_h =
\frac{\omega_h}{\sqrt{2E_\nu}}\left(
\begin{matrix}
\sqrt{E_\nu - hp_\nu} \\
\sqrt{E_\nu + hp_\nu}
\end{matrix}
\right) e^{-iE_\nu t + ip_\nu x} . 
\label{eq:nuhot}
\end{equation}
After decoupling the helicity is conserved during whole the evolution 
till the present epoch: $\nu \approx \nu_h \rightarrow \nu_h$ (see e.g. \cite{Long:2014zva}).  
Due to expansion of the Universe the energy and momentum 
redshift and at least two neutrinos become non-relativistic. For them the state
(\ref{eq:nuhot}) equals   
\begin{equation}
\nu_h =
\frac{\omega_h}{\sqrt{2}}\left(
\begin{matrix}
1 - hp_\nu/2m_\nu\\
1 + hp_\nu/2m_\nu      
\end{matrix}
\right) \approx 
\frac{\omega_h}{\sqrt{2}}\left(
\begin{matrix}
1 \\
1
\end{matrix}
\right) 
\label{eq:nucold}
\end{equation}
(where common phase factor is omitted).  Now both the left and right 
components have nearly equal amplitudes. That is,  the original $\nu_{Lh}$ state evolves into 
combination of the left and right states with equal weights. 
Indeed, in the non-relativistic limit  the left and right components are  
\begin{equation}
\nu_{Lh} =
\frac{\omega_h}{\sqrt{2}}\left(
\begin{matrix}
1 \\
0
\end{matrix}
\right),  ~~~~ \nu_{Rh} =
\frac{\omega_h}{\sqrt{2}}\left(
\begin{matrix}
0 \\
1
\end{matrix}
\right)   
\label{eq:nucoldlr}
\end{equation}
and their sum equals $(\nu_{Lh} +  \nu_{Rh}) = \nu_h$
(\ref{eq:nucold}). 
Thus, the   expansion of the Universe leads to transition 
$$
\nu_L \rightarrow (\nu_L + \nu_R) 
$$ 
as a result of adiabatic 
decrease of $p$ without oscillations. 

\section{Conclusion}

The claim of existence of the chiral oscillations in vacuum originates from   
usage of incorrect initial neutrino state. It is assumed that  neutrino produced in the chiral interactions is 
described by the left handed spinor $P_L u_h$ (or right handed spinor  $P_R u_h$),  
which corresponds to the Lorentz structure of Hamiltonian
of interactions. Expansion of such a spinor into eigenstates of propagation requires  eigenstates with 
both positive and negative energies. Thus,  the energy splitting $2E$ appears leading to oscillations with
frequency $2E_\nu$.  The same result can be obtained acting on $P_L u_h$ by the evolution matrix 
which steams from the Dirac equation. \\

We show that the state produced in the chiral interactions is not the chiral state described by 
$P_L u_h$. The produced state is a combination of eigenstates with different helicities but the same energy. 
Therefore no time dependent phase difference is generated,  and consequently, no oscillations
occur. Formally one can write any combination of solutions of the Dirac equation 
and they may have unusual properties. The point is that  it does not mean that they
are realized  in nature. \\

The produced neutrino state in a given process
should be computed rather than taken from the Lagrangian of interactions. 
Apart from the Loretz structure of interactions the state depends on kinematics of specific  production process. 
In this connection description of oscillations is elaborated in terms 
of amplitudes of production and interaction of neutrino eigenstates. 
Mismatch of the production and detection amplitudes leads to certain 
zero distance effect which does not depend on space-time. \\

For illustration some amplitudes were computed explicitly and it is shown 
that the ratio of amplitudes with different helicities 
contains not only expected $m_\nu/2E_\nu$ factor but also factors which depends on angles 
between neutrino and other particles involved. The latter dependence may disappear  
after integration of probabilities over the angles.  \\
   
Results for the Dirac and Majorana neutrinos are essentially 
the same with the only difference that in the Majorana case the zero distance 
effect leads to appearance of antineutrino which can produce positron, thus violating lepton number. 
In the case of several masses and flavor mixing the results 
are the same as in the case of pure flavor mixing.  In particular, no additional 
oscillation modes appear with 
frequency $2E_\nu$. Zero distance (time independent) effect may show up 
in the same way as in the case of single mass eigenstate. \\

In matter the neutrino components with different helicities have different interactions 
which leads to different matter potentials. Due to this the phase difference develops 
which produces oscillations. The frequency of oscillations is given by the difference of potentials,  
while the depths of oscillations is still  determined by the chiral suppression factor $(m_\nu/2E_\nu)^2$.  \\

There is no chiral oscillations of relic neutrinos. 
For the relic neutrino the adiabatic transition of $\nu_L$ to 
the state with equal amplitudes of $\nu_L$ and $\nu_R$ occurs.\\

Smallness of neutrino masses 
makes practically unobservable possible oscillation effects discussed here. 
In this connection one can consider their applications   
to heavy neutral leptons or usual charged fermions.

\section*{Appendix A.  ``The chiral oscillations'': some details}

In the papers on chiral oscillations it was assumed that neutrino state produced 
in the chiral interactions is described
by the chiral spinor defined as $u_{Lh} = 
P_L u_h$, where $P_L \equiv (1 - \gamma_5)/2$. 
It was further  assumed that this state  evolves according to the Dirac equation. 
Explicitly, for neutrino propagating in $z-$direction the Hamiltonian (\ref{eq:ham})
equals
\begin{equation}
H =
\left(
\begin{matrix}
- p_\nu \sigma_z & m_\nu I\\
m_\nu I & p_\nu \sigma_z
\end{matrix}
\right).
\label{eq:hamexpl}
\end{equation} 
The corresponding evolution matrix can be written as    
\begin{equation}
U (t) =  e^{-iHt}  = I \cos E_\nu t - \frac{i}{E_\nu} H \sin E_\nu t. 
\label{eq:evmat}
\end{equation} 
Notice that the oscillatory factors already appear  in this evolution matrix.  
However,  acting on correct (physical) states 
(in particular, states with definite helicities) they disappear. 
In this case the evolution is reduced to appearance of common phase factor.

In contrast, the evolution of the initial state (\ref{eq:evol}) produces
\begin{equation} 
\psi_{L h}(t) = \left [I \cos E_\nu t  - \frac{i}{E_\nu} H \sin E_\nu t 
\right] u_{L h}^{norm} = \left( \begin{matrix} [\cos E_\nu t + \sin E_\nu t ~\frac{ip_\nu }{E_\nu}
\sigma_z] \omega_h\\ 
- \sin E_\nu t ~\frac{im_\nu }{E_\nu } I \omega_h \end{matrix} \right). 
\label{eq:normev} 
\end{equation} 
Here we used that 
$\sigma_z \omega_h = h\omega_h$. According to (\ref{eq:normev}) 
the amplitude of probability to observe the initial state at the moment of time $t$  
equals  
\begin{equation} 
A_{LL} = \psi_{L h} (0)^{\dagger} \psi_{L h} (t) = \left[ \cos E_\nu t  + i \frac{h p_\nu}{E_\nu}
\sin E_\nu t \right], 
\label{eq:ampl}
\end{equation}
where the overal spatial factor $\exp({\bf p L})$ is omitted. 
This amplitude gives the survival probability ($L \rightarrow L$ transition)
\begin{equation}
P_{LL} = |\psi_{L h} (0)^{\dagger} \psi_{L h} (t)|^2 =  1 - \frac{m_\nu^2}{E_\nu^2} \sin^2
Et .   
\label{eq:prob}
\end{equation}
The probability of $L \rightarrow R$ transition, {\it i.e.},  the probability 
to observe $\psi_{R h} (0) = P_R \psi_{h}(0)$ equals the one in Eq. (\ref{eq:probLR}).  
These are the ``chiral oscillations'' with the eigenfrequency 
$\Delta H = 2E_\nu$ and the depth  $m_\nu^2/E_\nu^2$
(see, e.g.,  Ref.  \cite{Li:2023iys}).

\section*{Appendix B. Chiral oscillations and Zitterbewegung}

The chiral neutrino oscillations are to a large extent similar   
to the   Zitterbewegung 
(tremble motion) defined as a rapid ``oscillatory'' motion
of elementary particles  \cite{Bernardini:2007ew}, 
\cite{Eckstein:2016tnq}. Notion of 
Zitterbewegung is related to  interpretation of solution of the Dirac equation 
in terms of two  component spinors.  

Recall that the solution of the Dirac equation 
in the chiral basis is 
\begin{equation}
\psi  =
\left(
\begin{matrix}
\psi_L\\
\psi_R
\end{matrix}
\right), 
\label{eq:4spin}
\end{equation}
and in terms of $\psi_L$ and $\psi_R$ components the Dirac equation  
can be written as a system of two coupled equations for $\psi_L$ and $\psi_R$: 
\begin{equation}
i \slashed{\partial}\psi_L  =  m \psi_R, ~~~~~i \slashed{\partial}\psi_R  =  m
\psi_L.  
\label{eq:lrequat}
\end{equation}
According to (\ref{eq:lrequat}) $\psi_R$ turns out to be a source for $\psi_L$ and 
vice versa: $\psi_L$ is the source for $\psi_R$. 
Separately they do not satisfy the Dirac equation, 
while the  sum $\psi = \psi_L + \psi_R$ does. Furthermore, they are not the
eigenstates of Hamiltonian, e.g.,      
\begin{equation}
H \psi_{L h} = - hp \psi_{L h}  + m \frac{1}{\sqrt{2E}} 
\left(
\begin{matrix}
0 \\
\omega_-
\end{matrix}
\right), 
\label{eq:hpsil}
\end{equation}
and $- hp = E$. 
But they can be presented as combinations of solutions of the Dirac equation 
with positive and negative energies (or positive and negative time).

In the massless limit, $m = 0$,  $\psi_L$ and $\psi_R$ do satisfy the Dirac equation 
being the eigenstates of propagation. They appear as independent massless 
particles with definite and opposite helicities. 
This allows one to give  the following interpretation of the solution 
of the Dirac equation which implies phenomenon of Zitterbewegung. 
Namely,  massive Dirac particle can be considered  as a system of two massless
particles, which couple by the mass term.
The mass $m$ is treated as coupling constant and 
the coupling leads to {\it continuous}  transformations
of the  massless particles one into another:  
$\psi_L \leftrightarrow \psi_R$. The mass determines the rate
of transformations so that  the frequency of transformation  equals $2m$. 
The depth of transformation is proportional to $2m$.

As a result of transformation induced by mass the speed of the system is smaller than velocity of light. 
This is another interpretation of  effect of mass. 
To understand this phenomenon Penrose    \cite{penrose}  invoked the zigzag picture
which discretizes the process of transformations. 
In fact, it is this discretization that produces confusion. 
The propagation is described as series of transitions
$\psi_L \rightarrow \psi_R \rightarrow \psi_L \rightarrow ...$
Since the $\psi_L$, called ``zig'',  and  $\psi_R$, called ``zag'', have opposite
helicities but angular momentum is conserved, $\psi_R$
should move in opposite direction with respect to $\psi_L$.
So,  the function $x = x(t)$ has zig-zag form with respect to average $x = vt$.
This decreases effective velocity of the system.

The discrete zigzag picture (sometime the authors
call it fluctuations or oscillations) is just illustrative approximation.
Even if the pure state $\psi_L$ (or $\psi_R$) is produced
its transformation is continuous and not discrete.
Even in zigzag picture there is no fixed spatial
scale of zig and zag and one should consider all possible
lengths of zig and zag parts of trajectory and 
integrate over coordinates of vertices.

Effect of ``coupling" $m$ can be considered as effect of refraction (which is especially 
clear in the Standard Model) in vacuum ($hV$), that is, as  
appearance of refraction index which changes the velocity.
The key issue is  how to interpret zig-zag transitions.
In  \cite{penrose} no evolution equation is given  but the propagator of the system is discussed. 
The propagator is presented as 
superpositions of infinite number of quantum  transitions: sum of finite 
zigzags.  It is presented as matrix in the $\psi_L, \psi_R$ space with e.g. 
$\bar{\psi}_L \psi_L$ element equals the sum $\psi_L + (\psi_L \rightarrow \psi_R  
\rightarrow \psi_L) + 
(\psi_L  \rightarrow \psi_R   \rightarrow \psi_L  \rightarrow \psi_R   \rightarrow
\psi_L) + ... $
Actually, this presentation corresponds to well known computation of the propagator 
of massive fermion via propagators of massless fermions 
and mass as interaction vertex. Indeed, in the Lagrangian that leads to the Dirac
equation 
the mass terms equals 
$m \bar{\psi}_L \psi_R + h.c.$. The the propagator of massive particles is then the
infinite sum 
\begin{eqnarray}
& & \frac{\slashed{p}}{p^2}  + \frac{\slashed{ p}}{p^2}m\frac{\slashed{p}}{p^2} + 
\frac{\slashed{p}}{p^2}m\frac{\slashed{p}}{p^2} m \frac{\slashed{p}}{p^2}  + 
\frac{\slashed{p}}{p^2}m\frac{\slashed{p}}{p^2} m \frac{\slashed{p}}{p^2} m
\frac{\slashed{p}}{p^2} + ...
\nonumber\\
& = & \frac{\slashed{p}}{p^2}\left[1 + \frac{m^2}{p^2} + \left( \frac{m^2}{p^2}
\right)^2 + ...\right]  
+ \frac{m}{p^2} \left[1 + \frac{m^2}{p^2} + \left( \frac{m^2}{p^2} \right)^2 +
...\right] = 
\frac{\slashed{p} + m}{p^2 - m^2}, 
\label{eq:sumprop}
\end{eqnarray}
where the first sum in the second line follows from the odd terms of the first line, 
 while the second sum - from even terms of the first line. 
So,  we obtain the standard propagator and no reverse motion is needed.

\section*{Acknowledgments}

I should  acknowledge discussions with R. Gandhi in the middle of 90ies
after one of the first papers on chiral oscillations  appeared. 
Our preliminary conclusion  was that these oscillations  do not exist but we did not published our results. 
The author is grateful to E. Kh.  Akhmedov for joint work in the early stages of the project development. 
To a large extent the content of the paper was formed  
by intensive  discussions  with Georg Raffelt during more than half a year. 
I would like to thank Shao-Feng Ge for his comments and C.C. Nishi for useful communications.

\end{document}